\documentclass[conference,a4paper]{APSIPA2020}
\usepackage{multirow}
\usepackage{graphicx}
\usepackage{amsmath}
\usepackage[psamsfonts]{amssymb}
\usepackage{amsxtra}
\usepackage{threeparttable}
\usepackage{cite}
\usepackage{amsmath,amssymb,amsfonts}
\usepackage{graphicx}
\usepackage{subfig}
\usepackage{booktabs}
\usepackage{caption}
\usepackage{textcomp}
\usepackage{multirow}
\usepackage{tabularx}
\usepackage{xcolor}
\newcommand{\vect}[1]{\boldsymbol{#1}}
\usepackage[ruled,vlined]{algorithm2e}

\begin{document}

\title{Visual Security Evaluation of Learnable Image Encryption Methods against Ciphertext-only Attacks}

\author{%
\authorblockN{%
Warit Sirichotedumrong and
Hitoshi Kiya
}
\authorblockA{%
Department of Computer Science, Tokyo Metropolitan University, Hino-shi, Tokyo, Japan
}
}

\maketitle
\thispagestyle{empty}

\begin{abstract}
Various methods for protecting visual information have been proposed for privacy-preserving deep neural networks (DNNs). In contrast, methods of attack against such protection methods have been simultaneously developed. In this paper, we evaluate state-of-the-art visual-protection methods for privacy-preserving DNNs in terms of visual security against ciphertext-only attacks (COAs). We focus on the brute-force attack, feature reconstruction attack (FR-Attack), inverse transformation attack (ITN-Attack), and GAN-based attack (GAN-Attack), which have been proposed to reconstruct the visual information of plain images from visually protected images. The details of the various attacks are first summarized, and the visual security of the protection methods is then evaluated. Experimental results demonstrate that most of the protection methods, including pixel-wise encryption, are not robust enough against GAN-Attack, while a few are robust enough against it.
\end{abstract}

\section{Introduction}
Deep neural networks (DNNs) have greatly contributed to solving complex problems\cite{lecun2015deep,Donahue2014,Krizhevsky2012}, such as for computer vision, biomedical systems, natural language processing, and speech recognition. By utilizing a large amount of data to extract representations of relevant features, the performance has been greatly enhanced. Therefore, DNNs have been deployed in privacy-sensitive/security-critical applications, such as facial recognition, biometric authentication, and medical image analysis.

Recently, with the development of cloud services, DNNs are often carried in cloud environments. One of the advantages that cloud environments provide is that a large amount of data can be computed and processed by using cloud servers instead of using local servers. The other advantage is that cloud providers can provide various web-based software services like software as a service (SaaS). However, since cloud providers are semi-trusted, data privacy, such as personal information and medical records, may be compromised in cloud computing. Therefore, it is necessary to protect data privacy in cloud environments, so privacy-preserving DNNs have become an urgent challenge.

Various methods for protecting visual information have been proposed to protect the visual information of plain images\cite{kurihara2015pcs,Gaata2016,warit2018icme,tanaka2018iccetw,itier2019tcsvt,chuman2019tifs,warit2019apsipa_trans,warit2019icip,warit2019access,Ito_trans,Warit2020eusipco}. In contrast to information theory-based encryption (like RSA and AES), images generated by protection methods have pixel values and can be directly applied to some image processing algorithms. Some of the methods\cite{tanaka2018iccetw,warit2019icip,warit2019access,Ito_trans,Warit2020eusipco} have been proposed not only to protect the visual information of images but also for application to DNNs. In contrast, several ciphertext-only attacks have been proposed to reconstruct the visual information of plain images from encrypted ones simultaneously\cite{warit2019gcce,arxiv_coa,gan_attack}.

However, the visual security of state-of-the-art protection methods for privacy-preserving DNNs has not been extensively evaluated yet. This paper aims to evaluate the visual security of such methods against ciphertext-only attacks (COAs). In this paper, we focus on the brute-force attack, feature reconstruction attack (FR-Attack)\cite{arxiv_coa}, inverse transformation attack with pairs of plain and encrypted images (ITN-Attack\cite{warit2019gcce}, and GAN-based attack (GAN-Attack)\cite{gan_attack}, which have been proposed to reconstruct the visual information of plain images from visually protected images.

\section{Related Work}
\label{sec:related}
\subsection{Visual Information Protection}
\begin{figure}[!t]
\captionsetup[subfigure]{justification=centering}
\centering
\subfloat[]{\includegraphics[clip, height=2.5cm]{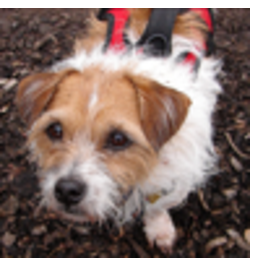}
\label{fig:label-A}}
\hfil
\subfloat[]{\includegraphics[clip, height=2.5cm]{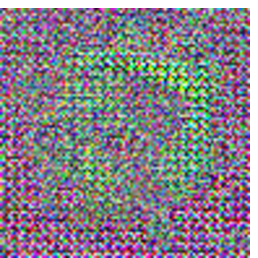}
\label{fig:label-B}}
\\
\subfloat[]{\includegraphics[clip, height=2.5cm]{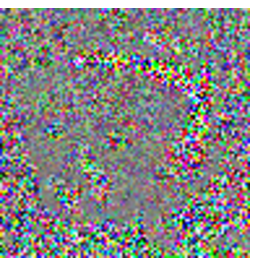}
\label{fig:label-C}}
\hfil
\subfloat[]{\includegraphics[clip, height=2.5cm]{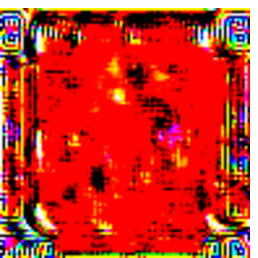}
\label{fig:label-D}}
\hfil
\subfloat[]{\includegraphics[clip, height=2.5cm]{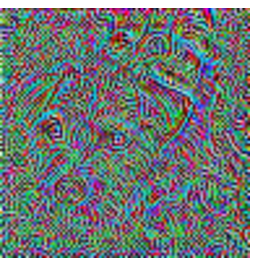}
\label{fig:label-E}}
\caption{Example of encrypted images. (a) Original image ($U \times V = 96 \times 96$). (b) Tanaka's scheme\cite{tanaka2018iccetw}. (c) Pixel-wise encryption\cite{warit2019icip,warit2019access}. (d) TN-model\cite{Ito_trans} (e) TN-GAN\cite{Warit2020eusipco}.}
\label{fig:eximages}
\end{figure}

Various perceptual image-encryption methods have been proposed to protect the visual information of plain images\cite{kurihara2015pcs,Gaata2016,warit2018icme,tanaka2018iccetw,itier2019tcsvt,chuman2019tifs,warit2019apsipa_trans,warit2019icip,warit2019access}. Visually protected images generated by using an encryption method consist of pixels, as shown in Figs.\,\ref{fig:eximages}(b) and \,\ref{fig:eximages}(c). Therefore, encrypted images can be directly applied to image processing algorithms. Some encryption methods have been studied for applying encrypted images to traditional machine learning algorithms, such as support vector machine (SVM), under the use of the kernel trick\cite{Kawamura2020ieice,Takahiro2019ieice}, but they cannot be applied to DNNs\cite{warit2019access}. There are four perceptual encryption methods\cite{tanaka2018iccetw,warit2019icip,warit2019access,Ito_trans} for privacy-preserving DNNs.

\subsection{Block-wise Image Encryption for DNNs}
A block-wise image encryption scheme was proposed for protecting the visual information of training and testing images for privacy-preserving DNNs\cite{tanaka2018iccetw}. This scheme is known as Tanaka's scheme\cite{tanaka2018iccetw}, which applies encrypted images to DNNs by adding an adaptation network prior to the DNNs for reducing the influence of image encryption.

\begin{figure*}[t]
\centering
\includegraphics[width=12cm]{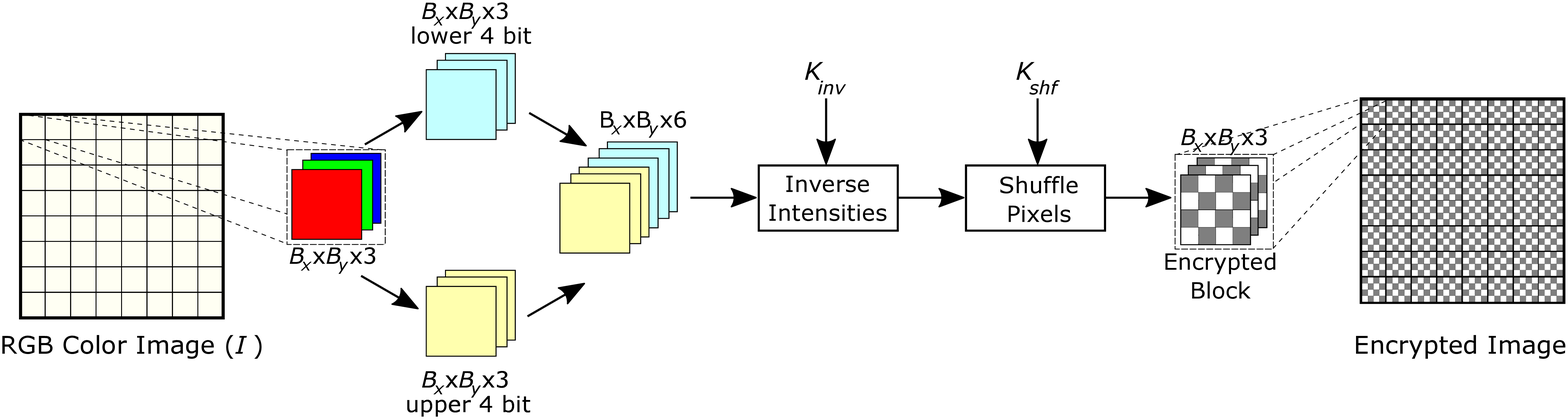}
\caption{Encryption process of Tanaka's scheme\cite{tanaka2018iccetw}, where $B_x\times B_y = 4\times 4$.}
\label{fig:enc_tanaka}
\end{figure*}

As depicted in Fig.\,\ref{fig:enc_tanaka}, 8-bit pixel values in $B_x\times B_y$ blocks are separated into upper and lower 4-bit pixel values to form 6-channel blocks, where $B_x\times B_y = 4\times 4$. The intensities of the pixel values in each block are randomly reversed by using secret key $K_{inv}$, and pixels in each block are then shuffled by using secret key $K_{shf}$, where $K_{inv}$ and $K_{shf}$ are applied to all blocks in common. The 6-channel blocks are reformed to 3-channel blocks to generate an encrypted image. 

The key space of Tanaka's method\cite{tanaka2018iccetw}, $N_{Tanaka}$, is given by
\begin{equation}
\label{eq:blocknum}
N_{Tanaka}=96!\cdot 2^{96}.
\end{equation}

In addition to using encrypted images, Tanaka's method utilizes an adaptation network, prior to using DNNs, to reduce the influence of image encryption. After the adaptation network, any network can follow.

\subsection{Pixel-wise Image Encryption for DNNs}

\begin{figure}[t]
\centering
\includegraphics[width =8cm]{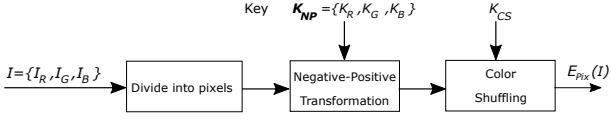}
\caption{Pixel-wise image encryption\cite{warit2019icip,warit2019access}}
\label{fig:enc_steps}
\end{figure}

A pixel-wise image encryption method was proposed for privacy-preserving DNNs that generates an encrypted image $E_{Pix}(I)$ from image $I$\cite{warit2019icip,warit2019access}. In this encryption method, the following steps are carried out (See Fig.\,\ref{fig:enc_steps}).
\begin{itemize}
\item [Step 1:] Divide RGB color image $I$ with $U \times V$ pixels into pixels.
\item [Step 2:] Individually apply negative-positive transformation (NP) to each pixel of three color channels, $I_{R}$, $I_{G}$, and $I_{B}$, by using a random binary integer generated by a set of secret keys $\vect{K_{NP}}=\{K_{R}, K_{G}, K_{B}\}$, where three keys, $K_{R}$, $K_{G}$, and $K_{B}$, are used for encrypting $I_{R}$, $I_{G}$, and $I_{B}$, respectively. 
\item [Step 3:] Shuffle three color components of each pixel by using an integer randomly selected from six integers generated by secret key $K_{CS}$.
\end{itemize}

Key set $\vect{K}=\{\vect{K_{NP}},K_{CS}\}$ is used for pixel-wise image encryption. There are two key conditions for encrypting $g$ training images $\vect{T}=\{I_{t_1}, I_{t_2}, \ldots, I_{t_g}\}$ and $h$ testing images $\vect{Q}=\{I_{q_1}, I_{q_2}, \ldots, I_{q_h}\}$.

\begin{itemize}
\item \textit{\textbf{Same encryption key:}} All training and testing images are encrypted by using only one secret key, i.e., \\$\vect{K_{t_1}}=\vect{K_{t_2}}=\ldots=\vect{K_{t_g}}=\vect{K_{q_1}}=\ldots=\vect{K_{q_h}}=\vect{K}$.
\item \textit{\textbf{Different encryption keys:}} Different secret keys are independently assigned to training and testing images, i.e., $\vect{K_{t_1}}\neq\vect{K_{t_2}}\neq\ldots\neq\vect{K_{t_g}}\neq\vect{K_{q_1}}\neq\ldots\neq\vect{K_{q_h}}$.
\end{itemize}
$\vect{K_{t_i}}$ is a key set used for encrypting training image $I_{t_i}$, and $\vect{K_{q_i}}$ is a key set for encrypting training image ${I_{q_i}}$.

Since all clients are able to utilize independent keys for training and testing a model, there is no need to manage the keys. 

If $I$ with $U \times V$ pixels is divided into pixels, the number of pixels $n$ is given by
\begin{equation}
\label{eq:blocknum}
n = U \times V.
\end{equation}
The key space of images encrypted by using pixel-wise encryption, $N_{Pix}(n)$, is represented by
\begin{equation}
N_{Pix}(n) = 2^{3n} \cdot 6^{n}.
\end{equation}
$N_{Pix}(n)$ is equal to $N_{Tanaka}$ when $n$ is approximately equal to 106.4. Therefore, pixel-wise encryption has a larger key space than Tanaka's method if $U \times V$ is more than $11 \times 11$ pixels.

\subsection{Image Transformation Network Trained with Model}
An image transformation network trained with a model (TN-model) was proposed that generates visually protected images for privacy-preserving DNNs\cite{Ito_trans}. In\cite{Ito_trans}, the transformation network allows us not only to generate visually protected images but also to maintain the performance of DNNs that using plain images has. However, a DNN model used for training the transformation network has to be trained by using plain images, prior to transformation network training. Namely, visually protected images cannot be applied to model training. An example of a visually protected image generated by the transformation network is illustrated in Fig.\,\ref{fig:eximages}(d).

\subsection{Image Transformation Network Trained with GAN}
An image transformation network trained with a generative adversarial network (TN-GAN)\cite{Warit2020eusipco} was proposed that protects the visual information of plain images for both training and testing images. This scheme trains an unpaired image-to-image translation using cycle-consistent adversarial networks (CycleGAN)\cite{zhu2017unpaired} to obtain a transformation network, which transforms plain images to visually protected ones. An example of a visually protected image generated by TN-GAN is shown in Fig.\,\ref{fig:eximages}(e).

\section{Visual Security Evaluation}
In this paper, we aim to evaluate the visual security of privacy-preserving DNNs in terms of robustness against various attacks including state-of-the-art ciphertext-only attacks (COAs). As previously described in Section\,\ref{sec:related}, a few privacy-preserving DNNs\cite{tanaka2018iccetw,warit2019icip,warit2019access} were confirmed to be robust against brute-force attacks. However, a state-of-the-art COA was proposed that reconstructs the visual information of plain images from visually protected ones\cite{arxiv_coa}. In this paper, we focus on four COAs: brute-force attack, feature reconstruction attack (FR-Attack)\cite{arxiv_coa}, GAN-based attack (GAN-Attack)\cite{gan_attack}, and inverse transformation network attack (ITN-Attack)\cite{warit2019gcce}. The above attacks are summarized below. 

\subsection{Brute-force Attack}
Block-wise encryption\cite{tanaka2018iccetw} and pixel-wise image encryption\cite{warit2019icip,warit2019access} carry out image transformation by using secret keys, so robustness against brute-force attacks has to be evaluated on the basis of key spaces. In contrast, the TN-GAN\cite{Warit2020eusipco} and TN-models\cite{Ito_trans} have no secret keys, so they are robust against brute-force attacks, although other COAs can be applied to images encrypted by using image transformation networks.

\subsection{Feature Reconstruction Attack}
A feature reconstruction attack (FR-Attack)\cite{arxiv_coa} has been proposed that reconstructs the edge information of plain images from encrypted images. To recover edge information, each pixel of an encrypted image is transformed by using Algorithm 1.

\begin{algorithm}
\caption{Feature reconstruction attack\cite{arxiv_coa}}
\SetKwInOut{Input}{Input}
\SetKwInOut{Output}{output}
\Input{Encrypted input image $I_e$ of size $U\times V$ ; number of bits $L$; leading bit $b\in\{0,1\}$.}%
\ForEach{$p=(u,v)\in I_e$}{%
\ForEach{$C\in \{R,G,B\}$}{%
\If{$\lfloor{{p_C}/(2L-1)}\rfloor\neq b$}%
{$p_C\gets p_C \oplus (2L-1)$\;}%
}%
}%
\end{algorithm}

\subsection{GAN-based Attack}

\begin{figure}[t]
\centering
\includegraphics[width =7.5cm]{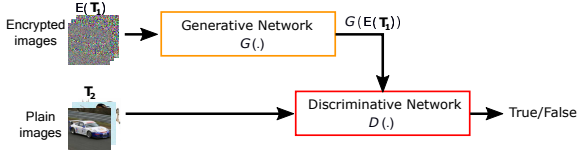}
\caption{Training framework of $G(.)$ and $D(.)$ for GAN-based attack}
\label{fig:dcgan}
\end{figure}

\begin{figure}[t]
\centering
\includegraphics[width =6.5cm]{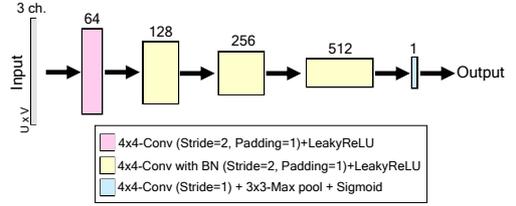}
\caption{Discriminative network $D(.)$ for GAN-based attack}
\label{fig:discriminator}
\end{figure} 
\begin{figure*}[t]
\centering
\includegraphics[width =15.5cm]{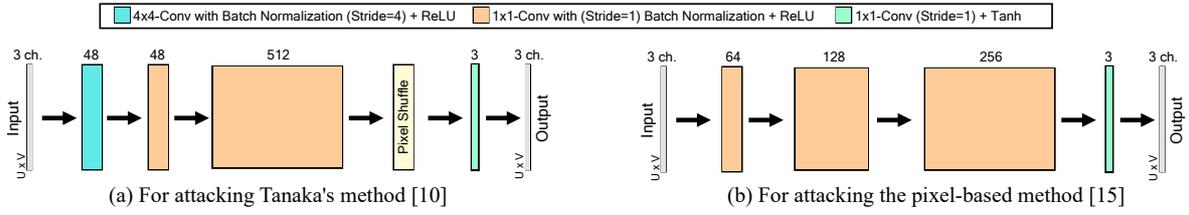}
\caption{Architectures of reconstruction network $G(.)$ for GAN-based attack}
\label{fig:generator}
\end{figure*}

Figure\,\ref{fig:dcgan} illustrates a training framework for a GAN-based attack (GAN-Attack)\cite{gan_attack} that consists of one generative network and one discriminative network. Generative network $G(.)$ reconstructs the visual information of plain images from encrypted images, while discriminative network $D(.)$ distinguishes the difference between reconstructed and plain images. The network architectures of $D(.)$ and $G(.)$ used in this paper are illustrated in Fig.\,\ref{fig:discriminator} and Fig.\,\ref{fig:generator}, respectively. 

To train $G(.)$ and $D(.)$, a set of training images $\vect{T}$ is equally divided into two image sets, $\vect{T_1}$ and $\vect{T_2}$, namely, $\vect{T_1}\neq\vect{T_2}$. Then, $\vect{T_1}$ is encrypted by using an encryption scheme ($E(.)$). Eventually, $E(\vect{T_1})$ and $\vect{T_2}$ are employed for training $G(.)$ and $D(.)$ (See Fig.\,\ref{fig:dcgan}). In the testing process, the reconstruction model $G(.)$, which is trained by $E(\vect{T_1})$, is utilized to recover encrypted test images ($E(\vect{Q})$) to obtain the reconstructed test images ($\vect{Q'}$).

This attack can be applied to visually protected images even when exact pairs of plain images and corresponding encrypted ones are not prepared.

\subsection{Inverse Transformation Network Attack}

\begin{figure}[t]
\centering
\includegraphics[width =8cm]{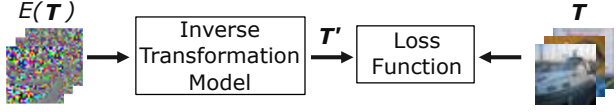}
\caption{Training framework of ITN-Attack, where $E(.)$ denotes encryption algorithm.}
\label{fig:pir_system}
\end{figure}

 \begin{figure}[t]
\centering
\includegraphics[width =7cm]{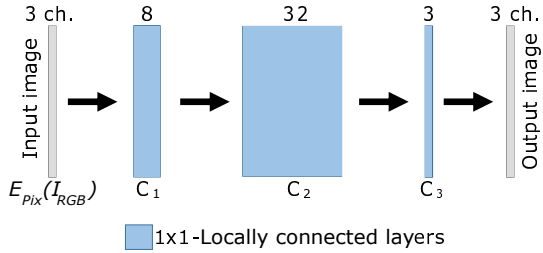}
\caption{Network architecture of reconstruction model. Each box denotes multi-channel feature map produced by each layer. Number of channels is denoted above each box. Feature map resolutions are $U \times V$. Kernel size and stride of locally connected layers are (1,1).}
\label{fig:pix_network}
\end{figure}

If adversaries can prepare exact pairs of plain images and corresponding encrypted ones, an inverse transformation network attack (ITN-Attack) can be applied to the protected images. Figure\,\ref{fig:pir_system} illustrates a training framework for ITN-Attack. An inverse transformation network is trained by using $E(\vect{T})$, and the training loss is then calculated from a set of reconstructed images ($\vect{T'}$) and $\vect{T}$ . Note that the network for ITN-Attack depends on the protection method. For example, Fig.\,\ref{fig:pix_network} depicts the architecture of ITN-Attack used for attacking pixel-wise image encryption, which consists of three 1$\times$1-locally connected layers ($C_1, C_2,$ and $C_3$) each with both a kernel size and a stride of (1,1). A locally connected layers similarly works as a 1$\times$1-convolution layer, but weights are unshared.

\section{Experiments}
We employed the STL-10 dataset, which consists of 5K training images and 8K testing images\cite{pmlr-v15-coates11a}, and each image has $96 \times 96$ pixels. Note that data augmentation and pre-processing were not applied to training images.

Robustness against attacks was evaluated in terms of the visibility of the reconstructed images. Moreover, we measured the average structural similarity (SSIM) values between the original and reconstructed images, where lower values mean less visual information. 

\subsection{Feature Reconstruction Attack}
\begin{figure*}[!t]
\captionsetup[subfigure]{justification=centering}
\centering
\subfloat[Tanaka's scheme]
{\includegraphics[clip, height=2.8cm]{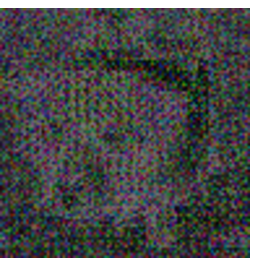}
\label{fig:label-A}}
\hfil
\subfloat[Pixel-wise (Same Key)]
{\includegraphics[clip, height=2.8cm]{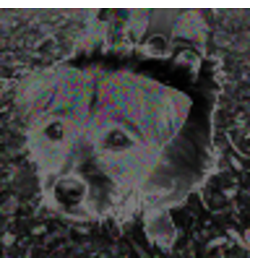}
\label{fig:label-b}}
\hfil
\subfloat[Pixel-wise (Different Key)]
{\includegraphics[clip, height=2.8cm]{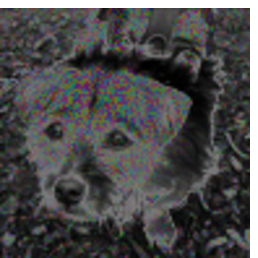}
\label{fig:label-B}}
\hfil
\subfloat[TN-model]
{\includegraphics[clip, height=2.8cm]{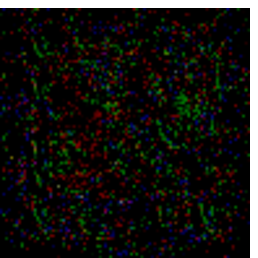}
\label{fig:label-D}}
\hfil
\subfloat[TN-GAN]
{\includegraphics[clip, height=2.8cm]{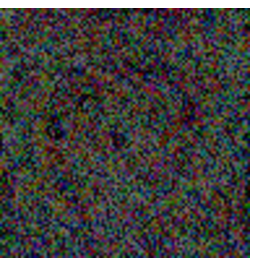}
\label{fig:label-E}}
\caption{Examples of images reconstructed by using FR-Attack\cite{arxiv_coa}}
\label{fig:recon_coa}
\end{figure*}

Figure\,\ref{fig:recon_coa} shows images reconstructed by using FR-Attack. For the pixel-wise method and Tanaka's scheme, the visual information of the plain image was recovered by FR-Attack, where Fig.\ref{fig:eximages}(a) is the plain one. In contrast, images encrypted by using the transformation networks were robust against FR-Attack. 

As shown in Table\,\ref{tbl:result}, TN-model provided the lowest SSIM value among the encryption schemes. Although TN-GAN offered almost the same SSIM value as that of Tanaka's scheme, the visibility of images encrypted by TN-GAN were lower than Tanaka's scheme. 

\begin{figure*}[!t]
\captionsetup[subfigure]{justification=centering}
\centering
\subfloat[Tanaka's scheme]
{\includegraphics[clip, height=2.8cm]{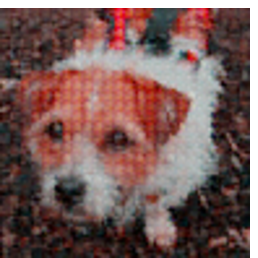}
\label{fig:label-A}}
\hfil
\subfloat[Pixel-wise (Same Key)]
{\includegraphics[clip, height=2.8cm]{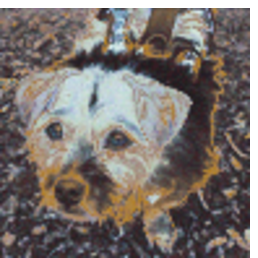}
\label{fig:label-B}}
\hfil
\subfloat[Pixel-wise (Different Key)]
{\includegraphics[clip, height=2.8cm]{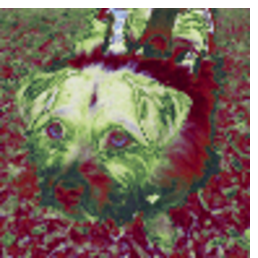}
\label{fig:label-C}}
\hfil
\subfloat[TN-model]
{\includegraphics[clip, height=2.8cm]{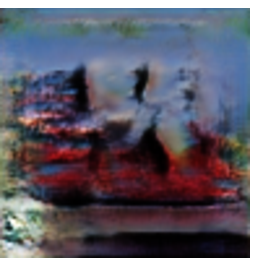}
\label{fig:label-D}}
\hfil
\subfloat[TN-GAN]
{\includegraphics[clip, height=2.8cm]{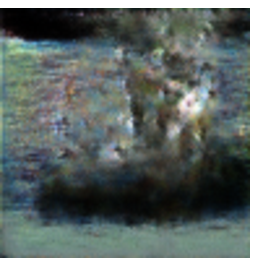}
\label{fig:label-E}}
\caption{Examples of images reconstructed by using GAN-Attack\cite{gan_attack}}
\label{fig:recon_gan}
\end{figure*}

\begin{figure*}[!t]
\captionsetup[subfigure]
{justification=centering}
\centering
\subfloat[Tanaka's scheme]
{\includegraphics[clip, height=2.8cm]{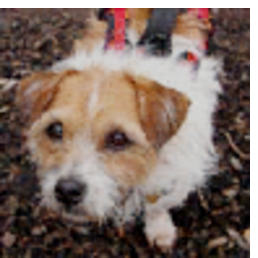}
\label{fig:label-A}}
\hfil
\subfloat[Pixel-wise (Same Key)]
{\includegraphics[clip, height=2.8cm]{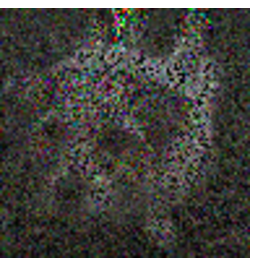}
\label{fig:label-B}}
\hfil
\subfloat[Pixel-wise (Different Key)]
{\includegraphics[clip, height=2.8cm]{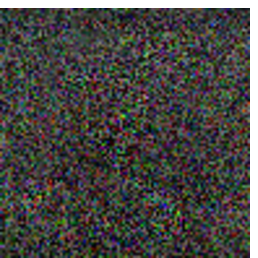}
\label{fig:label-C}}
\hfil
\subfloat[TN-model]
{\includegraphics[clip, height=2.8cm]{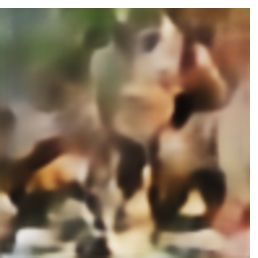}
\label{fig:label-D}}
\hfil
\subfloat[TN-GAN]
{\includegraphics[clip, height=2.8cm]{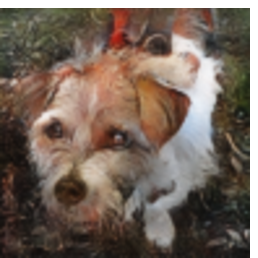}
\label{fig:label-E}}
\caption{Examples of images reconstructed by using ITN-Attack\cite{warit2019gcce}}
\label{fig:recon_pir}
\end{figure*}

\subsection{GAN-based Attack}
\subsubsection{Training Conditions}
GAN-Attack was trained for 100 epochs by using the Adam optimizer\cite{kingma2014adam} with a learning rate of 0.0002, a momentum parameter of ${\beta}=0.5$, and a batch size of 64. For reconstructing images encrypted by Tanaka's scheme and the pixel-wise method, we employed the network architectures in Figs.\,\ref{fig:generator}(a) and \,\ref{fig:generator}(b), respectively. U-Net\cite{unet} was utilized as $G(.)$ in order to reconstruct transformed images generated by TN-model and TN-GAN.

\subsubsection{Results}
Examples of images reconstructed by using GAN-Attack are shown in Fig.\,\ref{fig:recon_gan}, where Fig.\ref{fig:eximages}(a) is the original one. Although the visual information of the images encrypted by Tanaka's scheme and the pixel-wise one was reconstructed by GAN-Attack, it was difficult to reconstruct the visual information of the plain images from images generated by TN-model and TN-GAN.

As shown in Table\,\ref{tbl:result}, images generated by TN-model and TN-GAN had higher robustness against GAN-Attack than other methods. 

\subsection{Inverse Transformation Network Attack}
\subsubsection{Training Conditions}
For reconstructing images encrypted by the pixel-wise method, the network in Fig.\,\ref{fig:pix_network} was trained for 70 epochs by using stochastic gradient descent (SGD) with momentum for 70 epochs. The initial learning rate was set to 0.1 and was lowered by a factor of 10 at 40 and 60 epochs.

The network in Fig.\,\ref{fig:generator}(a) and U-Net\cite{unet} were trained by using SGD with momentum for 300 epochs to reconstruct images encrypted by Tanaka's scheme, TN-model, and TN-GAN. The learning rate was initially set to 0.1 and decreased by a factor of 10 at 150 and 225 epochs.

We used a weight decay of 0.0005, a momentum of 0.9, and a batch size of 128. The mean squared error (MSE) between $\vect{T'}$ and $\vect{T}$ was used as a loss function for all ITN-Attack experiments. 

\subsubsection{Results}
As demonstrated in Fig.\,\ref{fig:recon_pir}, images encrypted by using the pixel-wise method with different keys and TN-model were robust against ITN-Attack, while images encrypted by the other methods were reconstructed. Therefore, the pixel-wise method with different keys provided the lowest SSIM value.

Although TN-GAN is not robust against ITN-Attack, adversaries cannot prepare exact pairs of plain images and the corresponding protected ones because the weights of TN-GAN are not disclosed to the public.

\begin{table}
\centering
\caption{Average structural similarity (SSIM) values of images reconstructed by various attacks (N/A: not available)}
\label{tbl:result}
\begin{tabular}{cc|ccc}
\toprule
\multicolumn{2}{c|}{\multirow{3}{*}{\shortstack{Encryption\\ Scheme}}}&\multicolumn{3}{c}{SSIM values}\\
\cmidrule{3-5}
&&FR-Attack&ITN-Attack&GAN-Attack\\
&&\cite{arxiv_coa}&\cite{warit2019gcce}&\cite{gan_attack}\\
\midrule
Pixel-wise&Same&0.4646&0.1715&0.2688\\
\cmidrule{2-5}
\cite{warit2019icip,warit2019access}&Different&0.4628&0.0425&0.1527\\
\midrule
\multicolumn{2}{c|}{Tanaka's Scheme\cite{tanaka2018iccetw}}&0.1079&0.9147&0.7152\\
\midrule
\multicolumn{2}{c|}{TN-model\cite{Ito_trans}}&0.0263&0.3142&0.0793\\
\midrule
\multicolumn{2}{c|}{TN-GAN\cite{Warit2020eusipco}}&0.1093&0.7369&0.0956\\
\bottomrule
\end{tabular}
\end{table}

\begin{table*}
\centering
\caption{Properties of learnable image encryption in terms of robustness against various attacks, availability for model training/testing, and classification performance.}
\label{tbl:summary}
\begin{tabular}{cc|ccc|cc|cc}
\toprule
\multicolumn{2}{c|}{\multirow{3}{*}{\shortstack{Encryption\\ Scheme}}}&\multicolumn{3}{c|}{Attacks}&\multirow{2}{*}{\shortstack{Model Training}}&\multirow{2}{*}{\shortstack{Model Testing}}&\multirow{3}{*}{\shortstack{Classification\\ Performance}}&\multirow{3}{*}{\shortstack{Computational\\Cost}}\\
&&\multicolumn{3}{c|}{($\circ$: Robust, $\times$: Non-robust)}&&&&\\
\cmidrule{3-5}
&&FR-Attack\cite{arxiv_coa}&ITN-Attack\cite{warit2019gcce}&GAN-Attack\cite{gan_attack}&\multicolumn{2}{c|}{($\circ$: Available, $\times$: Not available)}&&\\
\midrule
Pixel-wise&Same&$\times$&$\times$&$\times$&$\circ$&$\circ$&High&Low\\
\cmidrule{2-9}
\cite{warit2019icip,warit2019access}&Different&$\times$&$\circ$&$\times$&$\circ$&$\circ$&High&Low\\
\midrule
\multicolumn{2}{c|}{Tanaka's Scheme\cite{tanaka2018iccetw}}&$\circ$&$\times$&$\times$&$\circ$&$\circ$&Low&Low\\
\midrule
\multicolumn{2}{c|}{TN-model\cite{Ito_trans}}&$\circ$&$\circ$&$\circ$&$\times$&$\circ$&High&High\\
\midrule
\multicolumn{2}{c|}{TN-GAN\cite{Warit2020eusipco}}&$\circ$&$\times$&$\circ$&$\circ$&$\circ$&High&High\\
\bottomrule
\end{tabular}
\end{table*}

\subsection{Summary of Evaluation}
Table\,\ref{tbl:summary} summarizes the properties of learnable image encryption methods in terms of robustness against various attacks, availability for model training/testing, classification performance, and computational cost. Since TN-model and TN-GAN employ DNNs to generate visually protected images, they have a higher computational cost than the other perceptual image-encryption methods. Although TN-model and TN-GAN have the same computational cost for image transformation, the cost for TN-GAN is higher than TN-model for training image transformation networks.

\section{Conclusion}
We compared methods for protecting visual information for privacy-preserving DNNs in terms of robustness against various COAs. In this paper, we focused on four attack methods: brute-force attack, feature reconstruction attack (FR-Attack), inverse transformation attack (ITN-Attack), and GAN-based attack (GAN-Attack). The experimental results demonstrated that two transformation networks, TN-model and TN-GAN, were robust enough, although they have a larger computational cost that the other methods. TN-model was most robust amongst the protection methods, but it cannot be applied to the protection of training images.

\end{document}